\documentclass[twocolumn,preprintnumbers,amsmath,amssymbm,prl]{revtex4}
\usepackage{epsfig}
\usepackage{graphicx}

\begin{document}
\title{Universal charge-mass relation: From black holes to atomic nuclei}
\author{Shahar Hod}
\address{The Ruppin Academic Center, Emeq Hefer 40250, Israel}
\address{ }
\address{The Hadassah Institute, Jerusalem 91010, Israel}
\date{\today}

\begin{abstract}
\ \ \ The cosmic censorship hypothesis, introduced by Penrose forty
years ago, is one of the corner stones of general relativity. This
conjecture asserts that spacetime singularities that arise in
gravitational collapse are always hidden inside of black holes. The
elimination of a black-hole horizon is ruled out by this principle
because that would expose naked singularities to distant observers.
We test the consistency of this prediction in a gedanken experiment
in which a charged object is swallowed by a charged black hole. We
find that the validity of the cosmic censorship conjecture requires
the existence of a charge-mass bound of the form
$q\leq\mu^{2/3}E^{-1/3}_c$, where $q$ and $\mu$ are the charge and
mass of the physical system respectively, and $E_c$ is the critical
electric field for pair-production. Applying this bound to charged
atomic nuclei, one finds an upper limit on the number $Z$ of protons
in a nucleus of given mass number $A$: $Z\leq
Z^*={\alpha}^{-1/3}A^{2/3}$, where $\alpha=e^2/\hbar$ is the fine
structure constant. We test the validity of this novel bound against
the $(Z,A)$-relation of atomic nuclei as deduced from the
Weizs\"acker semi-empirical mass formula.
\end{abstract}
\bigskip
\maketitle


%

What are the physical limitations on the magnitude of the electric
charge of a system characterized by general parameters such as size
and mass? In a purely classical context, one can construct a
quantity with dimensions of length from the mass and charge
parameters of the system: $R_c\equiv q^2/2\mu c^2$ \cite{Note1,Got}.
It was shown \cite{Lev,Erb} that classical electrodynamics is a
self-consistent theory only in describing the motion of charges with
a characteristic radius greater than the classical radius $R_c$. In
fact, all known physical systems are characterized by the relation
$R>R_c$, where $R$ is the circumscribing radius of the system. This
observation can be written as
\begin{equation}\label{Eq1}
q\leq (2\mu R)^{1/2}\  .
\end{equation}
(We shall use natural units in which $G=c=1$.)

Black holes, with their extreme gravitational binding, are the only
known objects in nature whose size can come close to the limit: an
extremal Reissner-Nordstr\"om black hole satisfies the relation
$R/R_c=2$ (other black holes satisfy $R/R_c>2$). On the other hand,
weakly self-gravitating systems conform to the bound (\ref{Eq1})
with orders of magnitude to spare. For example, atomic nuclei
satisfy the relation $R/R_c\sim 10^2-10^3$ and are therefore far
larger than their classical radius. Thus, even atomic nuclei, the
densest composite charged objects in nature (with negligible
self-gravity), conform to the bound (\ref{Eq1}) with more than an
order of magnitude to spare. This may suggest that for weakly
self-gravitating systems the bound (\ref{Eq1}) is a bit loose. This
observation immediately motivates one to look for a {\it stronger}
upper limit on the electric charge of a weakly self-gravitating
spherical object of given mass and radius.

In the following it will be shown that, the self-consistency of the
physics of black holes reveals the existence of a universal
charge-mass upper limit of the form:
\begin{equation}\label{Eq2}
q\leq\mu^{2/3}E^{-1/3}_c\  ,
\end{equation}
where $E_c$ is the critical electric field for pair-production
\cite{Emax1,Emax2}. This new bound would be stronger than bound
(\ref{Eq1}) for spherical objects with $\mu\leq 8R^3E^2_c$.

The influential theorems of Hawking and Penrose \cite{HawPen}
demonstrate that spacetime singularities are ubiquitous features of
general relativity, Einstein's theory of gravity. This implies that
general relativity itself predicts its own failure to describe the
physics of these extreme situations. Nevertheless, the utility of
general relativity in describing gravitational phenomena is
maintained by the cosmic censorship conjecture
\cite{Pen,Haw1,Brady}. The weak cosmic censorship conjecture (WCCC)
asserts that spacetime singularities that arise in gravitational
collapse are always hidden inside of black holes. This statement is
based on the common wisdom that singularities are not pervasive
\cite{Brady}.

The validity of the WCCC is a necessary condition to ensure the
predictability of the laws of physics \cite{Pen,Haw1,Brady}. The
conjecture, which is widely believed to be true, has become one of
the cornerstones of general relativity. Moreover, it is being
envisaged as a basic principle of nature. However, despite the
flurry of research over the years, the validity of this conjecture
is still an open question (see e.g.
\cite{Wald1,Sin,Clar,Vis,Price,Wald,His,KayWal,BekRos,Hub,QuiWal,Hod1,HodPir,Hod2,ForRom1,Hodever1,Hodever2,Har}
and references therein).

The destruction of a black-hole event horizon is ruled out by this
principle because it would expose the inner singularities to distant
observers. Moreover, the horizon area of a black hole,
$A_{\text{hor}}$, is associated with an entropy
$S_{BH}=A_{\text{hor}}/4\hbar$ \cite{Beken1}. Therefore, without any
obvious physical mechanism to compensate for the loss of the
black-hole enormous entropy, the destruction of the black-hole event
horizon would violate the generalized second law of thermodynamics
\cite{Beken1}. For these two reasons, any process which seems, at
first sight, to remove the black-hole horizon is expected to be
unphysical.

According to the uniqueness theorems \cite{un1,un2,un3,un4,un5}, all
stationary solutions of the Einstein-Maxwell equations are uniquely
described by the Kerr-Newman metric which is characterized by three
conserved parameters: the gravitational mass $M$, the angular
momentum $J$, and the electric charge $Q$. A black-hole solution
must satisfy the relation
\begin{equation}\label{Eq3}
M^2-Q^2-a^2 \geq 0\  ,
\end{equation}
where $a\equiv J/M$ is the specific angular momentum of the black
hole. Extreme black holes are the ones which saturate the relation
(\ref{Eq3}). As is well known, the Kerr-Newman metric with
$M^2-Q^2-a^2<0$ does not contain an event horizon, and it is
therefore associated with a naked singularity rather than a black
hole. In this work we inquire into the physical mechanism which
protects the black-hole horizon from being eliminated by the
absorption of charged objects which may ``supersaturate" the
extremality condition (\ref{Eq3}).

One of the earliest attempts to eliminate the horizon of a black
hole is due to Wald \cite{Wald}. Wald tried to "over-charge" an
extremal Reissner-Nordstr\"om black hole (characterized by Q = M) by
dropping into it a charged test particle whose charge-to-mass ratio
is larger than unity. Wald considered the specific case of a
particle which starts falling from spatial infinity (thus, the
particle's energy-at-infinity is larger than its rest mass). He has
shown that this attempt to over-charge the black hole would fail
because of the Coulomb potential barrier surrounding the black hole.

A more 'dangerous' version of Wald's original gedanken experiment is
one in which the charged particle is slowly lowered towards the
black hole. In this case, the energy delivered to the black hole
(the part contributed by the body's rest mass, see below) can be
red-shifted by letting the assimilation point approach the
black-hole horizon. On the other hand, the particle's charge is not
redshifted by the gravitational field of the black hole. At a first
sight the particle [characterized by a small (redshifted)
mass-energy] is not hindered from entering the black hole and
removing its horizon, thereby violating cosmic censorship.

Consider a charged body of rest mass $\mu$, electric charge $q$, and
proper radius $R$ approaching a charged Reissner-Nordstr\"om black
hole. (We assume $q>0$ without loss of generality). The
test-particle approximation imposes the constraint $\mu\ll R\ll M$.
This guarantees that the object has a negligible self-gravity and
that it is much smaller than the scale set by the black hole.

The external gravitational field of the Reissner-Nordstr\"om black
hole is given by
\begin{equation}\label{Eq4}
ds^2=-\Big(1-{{2M}\over{r}}+{{Q^2}\over{r^2}}\Big)dt^2+\Big(1-{{2M}\over{r}}+{{Q^2}\over{r^2}}\Big)^{-1}dr^2
+r^2d\Omega^2\  .
\end{equation}
The black hole's (event and inner) horizons are located at
$r_{\pm}=M\pm (M^2-Q^2)^{1/2}$.

The total energy ${\cal E}$ (energy-at-infinity) of the body in a
black-hole spacetime is made up of three contributions \cite{Hodst}:
\begin{itemize}
\item {${\cal E}_0 = \mu{(g_{00})}^{1/2}$, the energy associated
with the body's mass (red-shifted by the gravitational field).}
\item {${\cal E}_{\text{elec}} = qQ/r$, the electrostatic
interaction of the charged body with the external electric field.}
\item {${\cal E}_{\text{self}}$, the gravitationally induced
self-energy of the charged body. The third contribution, ${\cal
E}_{\text{self}}$, reflects the effect of the spacetime curvature on
the particle's electrostatic self-interaction. The physical origin
of this force is the distortion of the charge's long-range Coulomb
field by the spacetime curvature. This can also be interpreted as
being due to the image charge induced inside the (polarized) black
hole \cite{Linet,BekMay}. The self-interaction of a charged particle
in the black-hole spacetime results with a repulsive (i.e., directed
away from the black hole) self-force. A variety of techniques have
been used to demonstrate this effect in the black-hole spacetime. In
particular, one finds \cite{Zel,Loh} ${\cal
E}_{\text{self}}=Mq^2/2r^2$ in the Reissner-Nordstr\"om spacetime.}
\end{itemize}

Thus, the total energy of a charged body located at the radial
coordinate $r$ in the black-hole spacetime is given by
\begin{equation}\label{Eq5}
{\cal
E}(r)=\mu\Big(1-{{2M}\over{r}}+{{Q^2}\over{r^2}}\Big)^{1/2}+{{qQ}\over{r}}+{{Mq^2}\over{2r^2}}\
.
\end{equation}
The radial coordinate $r$ is related to the proper distance $\ell$
above the horizon through the relation
\begin{equation}\label{Eq6}
\ell(r)=\int_{r_+}^{r}(g_{rr})^{1/2}dr\  .
\end{equation}
Assuming $r-r_+\ll r_+-r_-$ (of course, this assumption can only be
valid for non-extremal black holes), one finds:
\begin{equation}\label{Eq7}
\ell(r)={{2r_+(r-r_+)^{1/2}}\over{(r_+-r_-)^{1/2}}}
\Big[1-{{r-r_+}\over{6(r_+-r_-)}}+O\Big({{r-r_+}\over{r_+}}\Big)\Big]\
.
\end{equation}
From (\ref{Eq7}) one obtains the inverse relation
\begin{equation}\label{Eq8}
r(\ell)=r_++(r_+-r_-){{\ell^2}\over{4r^2_+}}[1+O(\ell^2/r^2_+)]\ .
\end{equation}
Note that for extremal black holes Eq. (\ref{Eq6}) implies
$\ell=\infty$ for any point outside the horizon. Thus, Eq.
(\ref{Eq7}) is valid only for {\it non}-extremal black holes under
the assumption $r-r_+\ll r_+-r_-$. For non-extremal black holes this
amounts to the assumption $\ell\ll r_+$.

Taking cognizance of Eqs. (\ref{Eq5}) and (\ref{Eq8}), one finds
that the total energy of a charged particle at a proper distance
$\ell$ $(\ell\ll r_+)$ above the horizon of a non-extremal black
hole is given by:
\begin{equation}\label{Eq9}
{\cal
E}(\ell)={{\mu\ell(r_+-r_-)}\over{2r^2_+}}+{{qQ}\over{r_+}}-{{qQ\ell^2(r_+-r_-)}\over{4r^4_+}}+{{Mq^2}\over{2r^2_+}}\
.
\end{equation}
This expression is actually the effective potential governing the
motion of a charged body in the black-hole spacetime. Provided
$qQ>0$, it has a {\it maximal} height located at
$\ell=\ell^*(\mu,q;M,Q)=\mu r^2_+/qQ$.

The most challenging situation for the cosmic censorship conjecture
occurs when the energy-to-charge ratio of the captured particle is
as small as possible. This can be achieved if one slowly lowers the
body towards the black hole, providing it with the {\it minimal}
energy ${\cal E}_{\text{min}}={\cal E}(\ell^*)$ required in order to
overcome the potential barrier (recall that the effective potential
barrier has a maximum located at $\ell=\ell^*$). This is also true
for any charged object which is released to fall freely from
$\ell>\ell^*$ with the minimally required energy ${\cal E}(\ell^*)$.

The absorption of the charged object by the black hole results with
a change $\Delta M = {\cal E}(\ell^*)$ in the black-hole mass
(assuming that the energy delivered to the black hole is as small as
possible) and a change $\Delta Q=q$ in its charge. The condition for
the black hole to preserve its integrity after the assimilation of
the body is:
\begin{equation}\label{Eq10}
Q+q\leq M+{\cal E}(\ell^*)\  .
\end{equation}
Substituting ${\cal E}(\ell^*)$ from Eq. (\ref{Eq9}) one finds a
necessary and sufficient condition for removal of the black-hole
horizon:
\begin{equation}\label{Eq11}
(q-\epsilon)^2+{{2\epsilon}\over
M}\Big(\mu\ell^*-q^2-{{q{\ell^*}^2}\over{2M}}\Big)+{{q\epsilon^2}\over
M} <0  ,
\end{equation}
where $r_{\pm}\equiv M\pm\epsilon$. The expression on the l.h.s. of
(\ref{Eq11}) is minimized for $q=\epsilon+O(\epsilon^2/M)$, yielding
\begin{equation}\label{Eq12}
2\mu\ell^*-q^2-q{\ell^*}^2/M<0\  ,
\end{equation}
as a sufficient condition for elimination of the black-hole horizon.
Finally, substituting $\ell^*=\mu r^2_+/qQ$ into (\ref{Eq12}), one
finds
\begin{equation}\label{Eq13}
q^3>\mu^2/E_+\  ,
\end{equation}
as a sufficient condition for removal of the black-hole horizon,
where $E_+=Q/r^2_+=M^{-1}+O(\epsilon^2/M)$ is the black-hole
electric field in the vicinity of its horizon. An assimilation of a
charged object satisfying condition (\ref{Eq13}) by a charged black
hole would violate the cosmic censorship conjecture.

At this point it should be emphasized that Schwinger discharge of
the black hole (vacuum polarization effects) sets an upper bound on
the black-hole electric field \cite{Emax1,Emax2}:
\begin{equation}\label{Eq14}
E_+\leq E_c\equiv {{m^2_l}\over{e\hbar}}\  ,
\end{equation}
where $e$ is the elementary electric charge and $m_l$ is the rest
mass of the lightest stable charged particle. Thus, the validity of
the WCCC conjecture (namely, the integrity of the black-hole
horizon) requires the existence of a universal charge-mass bound of
the form
\begin{equation}\label{Eq15}
q\leq\mu^{2/3}E^{-1/3}_c\  .
\end{equation}
It is worth recalling that the test-particle approximation we have
used is valid for objects in the regime $\mu\ll R\ll E^{-1}_c$.
These inequalities are easily satisfied by charged objects with
$q\gg\mu$. The intriguing feature of our derivation is that it uses
a principle whose very meaning stems from gravitation (the cosmic
censorship principle) to derive a universal bound which has nothing
to do with gravitation [written out fully, the bound (\ref{Eq15})
would involve $\hbar$ and $c$, but not G]. This provides a striking
illustration of the unity of physics.

%
The lightest charged particle in nature is the electron, and one
should therefore take $m_l\to m_e$ in the bound (\ref{Eq15}). With
this value of $m_l$ it is straightforward to verify that atomic
nuclei conform to the upper bound (\ref{Eq15}). This in turn
guarantees that the absorption of charged nuclei by a black hole
would respect the cosmic censorship principle. Yet, we conjecture
that charged objects like atomic nuclei and quark nuggets whose size
is {\it smaller} than the Compton wavelength of the electron would
conform to an even tighter bound with $m_l\to m_p$, where $m_p$ is
the proton's rest mass. Below we shall discuss and test this
conjecture.

Consider a nucleus composed of $Z$ protons and $N$ neutrons. We
first point out that the Compton wavelength $\hbar/m_e$ of an
electron is much larger than the size of a typical nucleus $\sim
A^{1/3}\hbar/m_p$, where $A=Z+N$ is the baryon number (this is true
for all nuclei with mass numbers $A\lesssim 10^8$ \cite{Mad1}). This
fact implies that the wavefunction of an electron inside a nucleus
is almost identically zero. This in turn implies that the structure
of atomic nuclei is mainly determined by the physical properties of
the nucleons (protons and neutrons) that are trapped inside the
nucleus, whereas the electrons which are almost entirely left
outside the nucleus have almost no influence on its internal
structure.

Let us assume for a moment that we live in a world in which there
are no electrons. Since these light particles are not trapped inside
the nucleus, it is reasonable to expect that this assumption will
have no significant influence on the internal structure of dense
nuclear matter. To be precise, it is well known that electrons do
participate in nuclear radioactive processes (e.g., in the
beta-decay process), but since they are {\it not} trapped inside the
nucleus itself they have no significant influence on its internal
structure. (The internal structure itself is determined by the
protons and neutrons that compose the nuclei.) With this assumption,
the critical electric field is given by Eq. (\ref{Eq14}) with
$m_l\to m_p$.

The charge and mass of a nucleus are given by
\begin{equation}\label{Eq16}
q=Z|e|\ \ \ ; \ \ \ \mu=Zm_p+Nm_n-{\cal E}_B\simeq Am_p\  ,
\end{equation}
where ${\cal E}_B$ is the binding energy of the nucleus, which is
typically much smaller than its mass. Substituting Eq. (\ref{Eq16})
into the upper bound (\ref{Eq15}), one finds the $(Z,A)$-inequality:
\begin{equation}\label{Eq17}
Z\leq Z^*={\alpha}^{-1/3}A^{2/3}\  ,
\end{equation}
for charged matter of nuclear density, where $\alpha=e^2/\hbar\simeq
1/137$ is the fine structure constant.

The largest known completely stable nucleus is lead-208 which
contains 82 protons and 126 neutrons. This nucleus satisfies the
relation $Z/A^{2/3}\simeq 2.33$ and it therefore conforms to the
upper bound (\ref{Eq17}) by a factor of $\sim 2.2$. The largest
known artificially made nucleus contains $118$ protons and a total
number of $294$ nucleons \cite{Ogan}--- it satisfies the relation
$Z/A^{2/3}\simeq 2.67$ and it therefore conforms to the upper bound
(\ref{Eq17}) by a factor of $\sim 1.9$.

It is expected that even heavier meta-stable nuclei would be
produced in the forthcoming years using accelerator production
techniques. Some calculations suggest that nuclei of $A\sim 300$ to
$476$ with low excitation energies may exist for very long times
\cite{Nato}. Could these nuclei be able to threaten the validity of
the cosmic censorship conjecture by violating the $(Z,A)$-bound
(\ref{Eq17})? To answer this question, we shall investigate the
$(Z,A)$-relation of atomic nuclei as deduced from the well-known
semi-empirical mass formula \cite{Rohlf,Segre,Cook,Goo1}.

The binding energy ${\cal E}_B$ of a nucleus (that is, the
difference between its mass and the sum of the masses of its
individual constituents) is well approximated by the semi-empirical
mass formula, also known as Weizs\"acker's formula
\cite{Rohlf,Segre,Cook,Goo1}:
\begin{eqnarray}\label{Eq18}
{\cal E}_B(A,Z)=a_V A-a_s A^{2/3}-a_C
{{Z(Z-1)}\over{A^{1/3}}}\nonumber\\-a_A{{{(A-2Z)}^2}\over{A}}+{{a_P}\over{A^{1/2}}}\
.
\end{eqnarray}

This well-known formula is partially based on theory and partly on
empirical measurements. The theory is based on the liquid drop model
which treats the nucleus as a drop of incompressible nuclear fluid
composed of protons and neutrons (but {\it not} electrons!). The
five terms on the r.h.s of Eq. (\ref{Eq18}) correspond to the
cohesive binding of all the nucleons by the strong nuclear force,
the electrostatic mutual repulsion of the protons, a surface energy
term, an asymmetry term (derivable from the protons and neutrons
occupying independent quantum momentum states) and a pairing term
(partly derivable from the protons and neutrons occupying
independent quantum spin states) \cite{Rohlf,Segre,Cook,Goo1}. The
coefficients in the semi-empirical mass formula are calculated by
fitting to experimentally measured masses of nuclei.


By maximizing ${\cal E}_B(A,Z)$ with respect to $Z$, one finds the
number of protons of the most stable nucleus of atomic mass $A$:
\begin{equation}\label{Eq19}
Z(A)={A\over 2}{1\over{1+\beta A^{2/3}}}\ ,
\end{equation}
where $\beta\equiv {{a_C}\over{4a_A}}$. (For light nuclei this
expression reduces to the canonical relation $Z=A/2$.)

The requirement $Z(A)\leq Z^*(A)$ yields the quadratic equation
\begin{equation}\label{Eq20}
2\beta A^{2/3}-{(\alpha A)}^{1/3}+2\geq 0\  .
\end{equation}
The {\it experimentally} measured value of $\beta$ is $\sim
7.7\times 10^{-3}$ \cite{Rohlf,Segre,Cook,Goo1}. It is easy to
verify that with this value of $\beta$ the inequality (\ref{Eq20})
holds true for {\it all} $A$ values. It is worth emphasizing,
however, that as opposed to the loose bound (\ref{Eq1}) which is
respected by all nuclei with more than an order of magnitude to
spare, the new bound (\ref{Eq17}) is much stronger
--- $Z(A)$ is of the {\it same} order of magnitude
as the upper limit $Z^*(A)$. In fact, the ratio $Z(A)/Z^*(A)$
reaches a maximal value of $\simeq 0.56$.

Strange quark matter consisting of up, down, and strange quarks may
have an energy per baryon that is less than that of nuclear matter
\cite{Wit,Heis} and would then be the true ground state of baryonic
matter. The possible existence of metastable or even stable quark
nuggets (also known as strangelets) has been widely discussed
\cite{Bod1,Bod2,Bod3,Mad1}. It is believed that the net electric
charge of color-flavor locked strangelets \cite{Bod1,Bod2,Bod3,Mad1}
is concentrated near their surface. Thus, one expects a charge-mass
relation of the form $Z\propto A^{2/3}$. The upper bound
(\ref{Eq17}) limits the allowed value of the proportionality
coefficient to be less than $\alpha^{-1/3}$. The accepted estimate
for this constant is $\sim 0.3$ \cite{Mad1}, which indeed conform to
the upper limit (\ref{Eq17}).

In summary, an application of the cosmic censorship principle to a
gedanken experiment in which a charged object falls into a black
hole, enables us to reveal a universal relation between the maximal
electric charge and mass of any spherically symmetric object with
negligible self gravity: $q\leq\mu^{2/3}E^{-1/3}_c$. For objects
with nuclear matter density the upper bound corresponds to $Z\leq
Z^*={\alpha}^{-1/3}A^{2/3}$. This relation limits the charges of
objects such as atomic nuclei and quark nuggets. For these objects,
the new bound is more restrictive than other limits existing in the
literature.

\bigskip
\noindent
{\bf ACKNOWLEDGMENTS}
\bigskip

This research is supported by the Meltzer Science Foundation. I
thank Yael Oren and Arbel M. Ongo for helpful discussions.


\end{document}